\documentclass{PoS}

\title{Top-quark production at the LHC:
differential cross section and phenomenological applications.}

\ShortTitle{Top-quark production at the LHC}

\author{\speaker{Marco Guzzi},$^a$ Katerina Lipka$^{a}$ 
and Sven-Olaf Moch $^{b,c}$ \\
\llap{$^a$}Deutsches Elektronen-Synchrotron DESY\\
Notkestrasse 85, 22607 Hamburg, Germany\\
\llap{$^b$}Institut f\"ur Theoretische Physik, Universit\"at Hamburg, Luruper Chaussee 149
Hamburg, D-22761, Germany\\
\llap{$^c$}Deutsches Elektronen-Synchrotron DESY\\
Platanenallee 6, D-15738 Zeuthen, Germany\\
E-mail: \email{marco.guzzi@desy.de},\email{katerina.lipka@desy.de}, 
\email{sven-olaf.moch@desy.de}}

\abstract{
We discuss top-quark pair production at hadron colliders and 
review available calculations of differential top-pair production cross section in perturbative 
QCD at approximate next-to-next-to-leading order (NNLO) within the threshold resummation formalism.
These calculations are implemented into an open source program under development. 
We present phenomenological studies at the LHC that include transverse 
momentum and rapidity distribution of the top quarks at a center-of-mass energy of 7 TeV.
Preliminary results obtained with this program are in very good agreement with the recent LHC measurements.}
 
\FullConference{XXI International Workshop on Deep-Inelastic Scattering and Related Subjects\\
                 22-26 April 2013, Marseilles, France\\ 

                 {\tt Preprint DESY 13-141} }

\begin{document}

{\bf Introduction.}
Top-quark phenomenology at hadron colliders will be prominent in the near future 
and a new realm of precision calculations has been reached in order
to analyze the large inflow of data from LHC with unprecedented accuracy.
Precise measurements of top-quark pair total and differential cross sections at a center-of-mass energy $\sqrt{s}=7$ TeV  
are already published by the CMS \cite{Chatrchyan:2012saa} and ATLAS \cite{Aad:2012hg} 
collaborations and data at 8 TeV are publicly available.
The top-quark's ``on shell'' mass $m_t$ has been recently determined \cite{Chatrchyan:2013haa} within  
few GeV uncertainty and, for the first time, the strong coupling constant $\alpha_s(M_Z)$ is extracted by using  $t\bar{t}$ events.
The cross section for top-quark pair production gives us the possibility 
of investigating the correlation between parton distribution functions (PDFs) of the proton (in particular the gluon), 
top-quark mass $m_t$, and $\alpha_s$. In fact, about 80\% of the total cross section is ascribed to the 
gluon-gluon channel at the LHC at $\sqrt{s}=7$ TeV. Therefore, top-quark pair production is a standard candle to test 
QCD factorization, properties of the Standard Model (SM), 
and also to investigate physics beyond the SM. 
Studies are going on in order to establish the role of quantum corrections 
to the top mass and the vacuum stability conditions which are 
driven by the precise value of the mass of the Higgs boson \cite{Alekhin:2012py,Bezrukov:2012sa}. 
Furthermore, new observables to measure the top-quark mass at hadron colliders have been recently proposed \cite{Alioli:2013mxa}.
Therefore, there is a clear demand of precise theoretical predictions at highest possible order for comparisons with the data, 
in which systematic uncertainties associated with renormalization/factorization ($\mu_R,\mu_F$) and other scales 
are reduced by including higher orders in perturbative calculations.
The computation of the $t\bar{t}$ production cross section is a big challenge and 
required continuous efforts by the QCD community for reaching the state-of-art
of QCD radiative corrections and the development of calculational tools. 
QCD corrections to heavy quark production at colliders in the next-to-leading order (NLO) approximation are known since many years \cite{Nason:1987xz,Beenakker:1988bq,Beenakker:1990maa}, but the full NNLO 
$O(\alpha_s^4)$ inclusive cross section has been computed only recently \cite{Czakon:2013goa,Czakon:2012pz}.
Available exact NLO calculations for $t\bar{t}$ total and differential cross sections have been 
implemented into Monte Carlo numerical codes such as \textsc{MCFM} \cite{Campbell:2000bg},
\textsc{MC@NLO} \cite{Frixione:2003ei}, \textsc{POWHEG}  \cite{Alioli:2010xa}, 
\textsc{MadGraph/MadEvent} \cite{Alwall:2007st,Frederix:2009yq} while
the state-of-the-art QCD computation for the $t\bar{t}$ production total cross 
section at NNLO is implemented in the \textsc{C++} computer programs \textsc{Top++} \cite{Czakon:2013goa} 
and \textsc{Hathor} \cite{Aliev:2010zk}.

{\bf Resummation.} 
In the kinematic region near the partonic threshold of the $t\bar{t}$ pair, 
remnants of soft-gluon dynamics in hard scattering functions can be large and dominate high-order corrections.
These enhancements have logarithmic structures which are resummed by threshold resummation techniques 
\cite{Bonciani:1998vc,Kidonakis:1997gm} that organize double-logarithmic 
corrections to all orders and extend the predictive power of QCD to these kinematic regions.
In the past years, many efforts have been put into developments of approximate NNLO QCD calculations that include threshold resummation 
to assess $t\bar{t}$ total and differential cross section \cite{Kidonakis:2010dk,Moch:2008qy}.
Novel techniques in soft-collinear effective theory (SCET) have been also recently developed 
\cite{Ahrens:2011mw} and some of the differences 
between SCET and the traditional resummation approach are outlined in \cite{Kidonakis:2011ca}.
Threshold resummation can be derived in two different kinematic 
domains in which we have different kinds of logarithms that have to be resummed. 
These two kinematic regimes give the same prediction when the $t\bar{t}$ pair is produced at the mass threshold.
In the single-particle inclusive (1PI) kinematic, heavy quark 
hadroproduction is dominated by the partonic subprocess
$i(k_1) + j(k_2)\rightarrow Q(p_1) + X[\bar{Q}](p'_2)$, where $k_{1,2}$ are the incoming momenta of partons $(i,j)$, 
$p'_2=\bar{p}_2 + k $, $k$ is any additional radiation, and  $s_4=p'^2_2-m_t^2\rightarrow 0$ is the momentum at the threshold.
The pair-invariant mass (PIM) kinematic is defined by the reaction 
$i(k_1) + j(k_2)\rightarrow Q\bar{Q}(p') +X'(k)$, where the threshold $p'^2 = M^2 $ is reached when $X'(k)=0$.
The approximate NNLO differential cross section at parton level in 1PI kinematic can be written in a compact form as
\begin{eqnarray}
&&s^2\frac{\hat{\sigma}_{ij}}{du_1 dt_1}\vert_{1PI}=F^{Born}_{ij}
\frac{\alpha_s^2(\mu_R^2)}{\pi^2}\left\{D^{(3)}_{ij}\left[\frac{\ln^3(s_4/m_t^2)}{s_4}\right]_+ 
+ D^{(2)}_{ij}\left[\frac{\ln^2(s_4/m_t^2)}{s_4}\right]_+ 
\right.\nonumber\\
&&\left.
+ D^{(1)}_{ij}\left[\frac{\ln(s_4/m_t^2)}{s_4}\right]_+ 
+ D^{(0)}_{ij}\left[\frac{1}{s_4}\right]_+ 
+ R_{ij}\delta(s_4)\right\}\nonumber
\end{eqnarray}
where $s_4=s+t_1+u_1$, $t_1=(k_2-p_1)^2-m_t^2$, $u_1=(k_1-p_1)^2-m_t^2$, $\mu_R$ is the renormalization scale, 
and $D^{(k)}_{ij}$ and $R_{ij}$ (which depend on Mandelstam invariants $s$, $t_1$, and $u_1$) 
are defined by soft and hard structures of partonic reactions. 

{\bf Phenomenological applications.} 
Motivated by the rapid inflow of very precise 
LHC data of $t\bar{t}$ differential cross sections, 
we illustrate phenomenological applications obtained by 
our program which is under development and designed for calculations 
of total and differential cross sections for heavy-quark pair production  
at hadron colliders in both 1PI and PIM kinematics. 
The package incorporates an approximate NNLO QCD 
calculation within the threshold resummation formalism (details in \cite{GLM}).
Our goal is to create a flexible open source code for phenomenological 
analyses and quantitative comparisons of theory with data.
The cross section at hadronic level contains several input parameters such as $m_t$, $\mu_R$, $\mu_F$, 
PDFs, and $\alpha_s(M_Z)$, which can be varied to explore the sensitivity
of the theory prediction and estimate systematic uncertainties. 
This is of particular interest in a global fit analysis of PDFs in which one studies 
the impact of the variation of these parameters on the extracted PDFs and 
in particular the correlation of the gluon PDF with $\alpha_s(M_Z)$ and top mass.
Measurements of differential cross sections of top-quark pair production
have potential sensitivity to determine the gluon distribution in the proton 
\cite{Czakon:2013tha}, but the correlation between PDFs, $m_t$ and $\alpha_s$ 
has to be carefully taken into account. 
Simultaneous determination of these QCD parameters using top-pair production 
measurements at the LHC requires in particular fast computing tools for the calculation of  
top-pair differential cross sections at highest available order in QCD.

\begin{figure}[ht]
\begin{centering}
\includegraphics[width=0.41\textwidth, angle=-90]{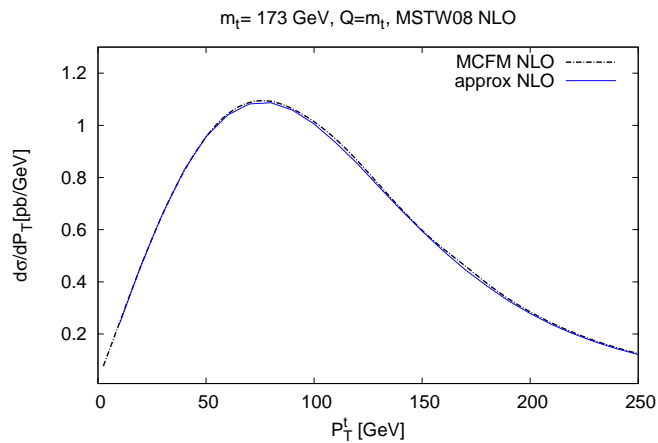} 
\par\end{centering}
\caption{\label{Qcheck} Exact NLO calculation obtained by using MCFM vs. approximate NLO.}
\end{figure}

\begin{figure}[ht]
\begin{centering}
\includegraphics[width=0.61\textwidth]{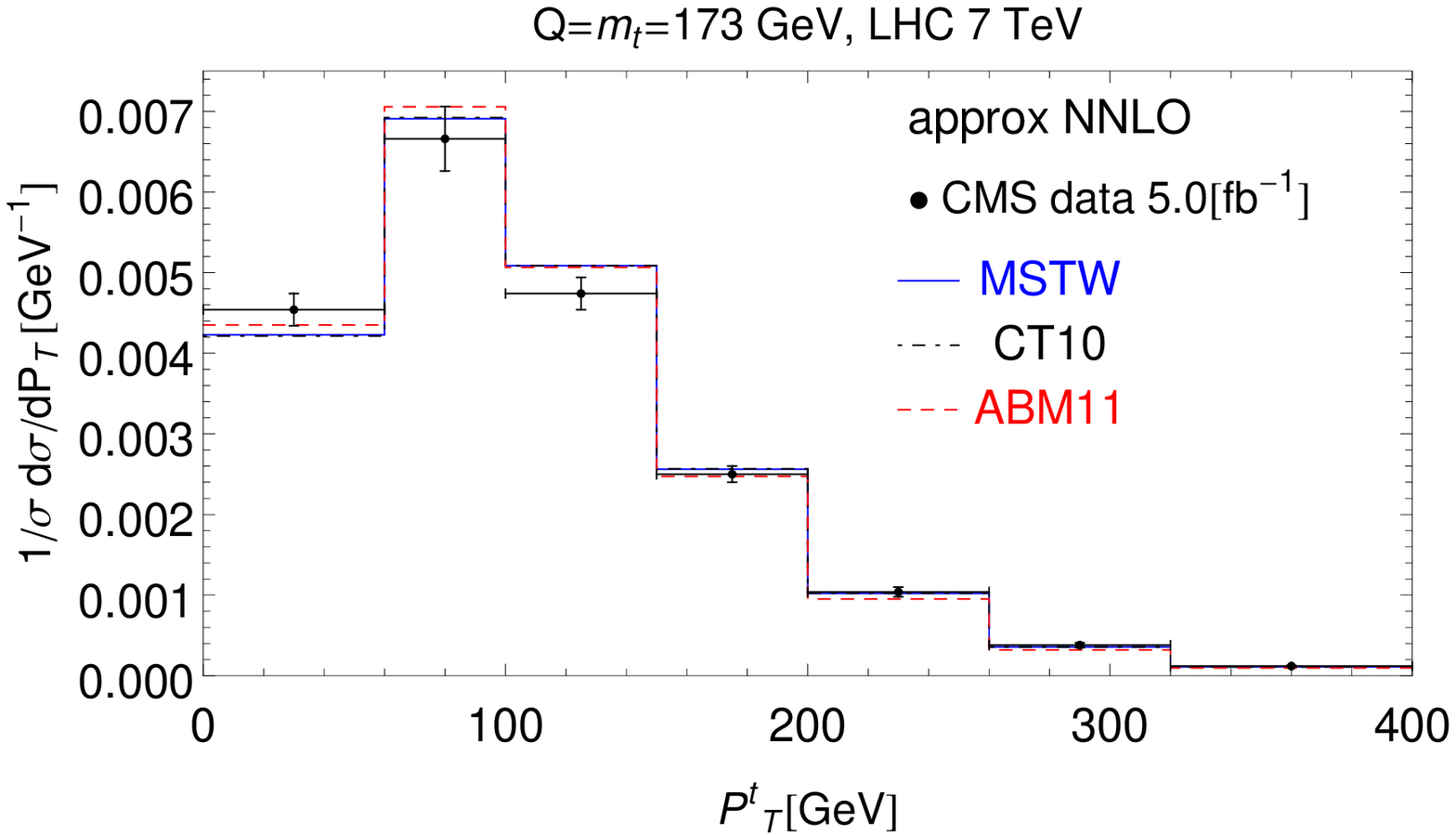}\\
\includegraphics[width=0.61\textwidth]{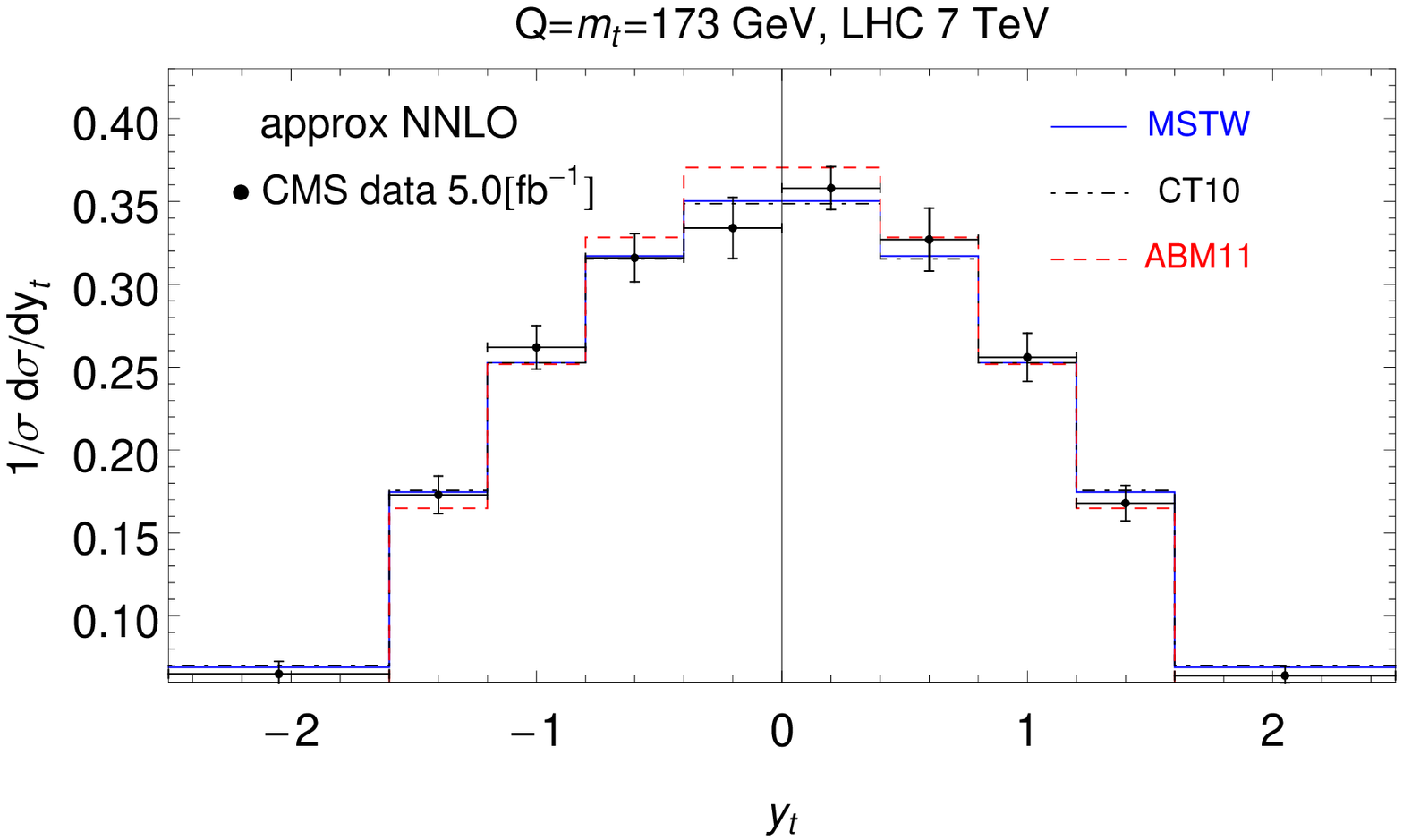} 
\par\end{centering}
\caption{\label{pty} Upper inset: normalized differential $t\bar{t}$ production 
cross section as a function of the transverse momentum of the top quarks.
Lower inset: same as upper inset but as a function of the rapidity distribution.}
\end{figure}

{\bf Results.} In Fig.~\ref{Qcheck} we show a comparison between the exact NLO calculation (black dot-dashed line) 
of the top-quark differential $P^{t}_T$ distribution at the LHC $\sqrt{s}=7$ TeV, 
obtained by using the MCFM code \cite{Campbell:2000bg} with MSTW \cite{Martin:2009iq} NLO PDFs 
and the approximate NLO calculation based on the threshold logarithms (solid), obtained with our code.
The default scale choice which we adopt here is $\mu_R=\mu_F=m_t$ and 
the value of the pole mass is set to $m_t=173$ GeV \cite{Kidonakis:2010dk}.
The excellent agreement shown in the  $P^{t}_T$ range around the threshold region 
is an indication of the fact that threshold resummation, supplemented with NLO matching conditions for hard and soft functions,
controls the large logarithms which dominate the cross section in this kinematic range 
and gives a very good estimate of the $P^{t}_T$ spectrum. Differences are below the percent level and cannot be resolved by 
the accuracy of the current data. Similar features are found in the rapidity distribution of top quarks.
In Fig.~\ref{pty} we compared the normalized approximate NNLO predictions 
for top-quark transverse momentum and rapidity ($y_t$) distributions 
to the recent measurements~\cite{Chatrchyan:2012saa} of the CMS experiment.
Predictions obtained by using different PDF sets are represented by histograms in which we show
MSTW~\cite{Martin:2009iq} (solid), CT10~\cite{Gao:2013xoa} (dot-dashed), 
and ABM11~\cite{Alekhin:2012ig} (dashed). 
The shape of both $P^t_T$ and $y_t$ differential distributions 
is well reproduced and the agreement to data is generally good for all PDFs where 
the recommended value of $\alpha_s(M_Z)$ is adopted in each case.
A spread in the theory predictions, more visible in the rapidity distribution, 
can be explained by different heavy quark treatments in the PDFs, different $\alpha_s(M_Z)$ values, 
and different preferred values (generally smaller in the ABM11 case) of the top-quark's pole mass.

{\bf Conclusions.}
The development of the open-source computing program at approximate 
NNLO for calculation of differential distributions of top-pair production 
in proton-(anti)proton collisions has started.
We have shown phenomenological studies at the LHC in which 
$t\bar{t}$ pair production differential cross sections are compared to recent CMS data. 
The NNLO approximate predictions are in overall good agreement with the data.

{\bf Acknowledgments.} This work was supported by the ``Initiative and Networking Fund of 
the Helmholtz Association (HGF) under the contract S0-072''.

\bibliographystyle{h-elsevier3}

\end{document}